\documentclass[sigconf,noacm]{acmart}
\renewcommand\footnotetextcopyrightpermission[1]{} 
\AtBeginDocument{%
  }

\usepackage{enumitem}
\usepackage{algorithmic}
\usepackage[ruled,linesnumbered, vlined]{algorithm2e}
\usepackage{ulem}
\usepackage{balance}
\setcopyright{acmlicensed}
\copyrightyear{2018}
\acmYear{2018}
\acmDOI{XXXXXXX.XXXXXXX}
\acmConference[ICPP '25]{}{September 8-11,
  2025}{San Diego, CA}
\acmISBN{978-1-4503-XXXX-X/2018/06}

\begin{document}

\title{TD-Pipe: \underline{T}emporally-\underline{D}isaggregated  Pipeline Parallelism Architecture for High-Throughput LLM Inference}

\author{Hongbin Zhang}
\email{zhanghb55@mail2.sysu.edu.cn}
\affiliation{%
  \institution{Sun Yat-sen University}
  \city{Guangzhou}
  \country{China}
}
\author{Taosheng Wei}
\email{weitsh@mail2.sysu.edu.cn}
\affiliation{%
  \institution{Sun Yat-sen University}
  \city{Guangzhou}
  \country{China}
}
\author{Zhenyi Zheng}
\email{zhengzhy37@mail2.sysu.edu.cn}
\affiliation{%
  \institution{Sun Yat-sen University}
  \city{Guangzhou}
  \country{China}
}
\author{Jiangsu Du}
\email{dujiangsu@mail.sysu.edu.cn}
\affiliation{%
  \institution{Sun Yat-sen University}
  \city{Guangzhou}
  \country{China}
}
\author{Zhiguang Chen}
\email{chenzhg29@mail.sysu.edu.cn}
\affiliation{%
  \institution{Sun Yat-sen University}
  \city{Guangzhou}
  \country{China}
}
\author{Yutong Lu}
\email{luyutong@mail.sysu.edu.cn}
\affiliation{%
  \institution{Sun Yat-sen University}
  \city{Guangzhou}
  \country{China}
}

\renewcommand{\shortauthors}{Hongbin Zhang et al.}
\begin{abstract}
As the model size continuously increases, 
pipeline parallelism shows great promise in throughput-oriented LLM inference due to its low demand on communications.
However, imbalanced pipeline workloads and complex data dependencies in the prefill and decode phases result in massive pipeline bubbles and further severe performance reduction.

To better exploit the pipeline parallelism for high-throughput LLM inference, we propose TD-Pipe, with the key idea lies in the temporally-disaggregated pipeline parallelism architecture.
Specifically, this architecture disaggregates the prefill and decode phases in the temporal dimension, so as to eliminate pipeline bubbles caused by the phase switching.
TD-Pipe identifies potential issues of exploiting the novel architecture and provides solutions.
First, a hierarchy-controller structure is used to better coordinate devices in pipeline parallelism by decoupling the scheduling from execution.
Second, the AI-based greedy prefill approach aggressively performs more prefills by predicting the output length and simulating the memory usage.
Third, the inter-batch work stealing approach dynamically balances decode phase workloads between different batches to reduce bubbles.
Forth, the spatial-temporal intensity comparison approach determines the optimal switch from decode to prefill by comparing the performance drop from reduced computational intensity with that from phase switching bubbles.
Extensive experiments show that TD-Pipe effectively increases the throughput of LLM inference by up to 1.91$\times$ over the existing tensor parallel approach and 2.73$\times$ over the existing pipeline parallel approach on GPU nodes with only PCIe interconnection.
\end{abstract}

\maketitle

\section{Introduction}
\label{sec:intro}

Large language models (LLMs) ~\cite{brown2020language, touvron2023llama, zeng2022glm, zhang2022opt, touvron2023llama2} demonstrate significant capabilities.
Besides delivering online LLM services~\cite{cai, chatgpt, AmazonCode, GithubCopilot}, there remain important scenarios where strict latency SLO constraints are unnecessary, such as batch APIs~\cite{openai-batch-api, anyscale-batch-api} and RLHF rollout stage~\cite{hu2024openrlhfeasytousescalablehighperformance}.
In these offline settings, maximizing the throughput has become the top priority.
However, given the substantial memory demands of LLMs and a single GPU's capacity is often insufficient, existing approaches~\cite{sheng2023flexgen, 2022deepspeed, shoeybi2019megatron, huang2019gpipe, dudu2024liger} cannot effectively coordinate multiple resources with minimal synchronization overhead for high-throughput LLM inference.

LLM generative inference entails substantial memory requirements and requires different resources to cooperate.
On the one hand, LLMs are renowned for their massive weights.
The number of parameters can be 7B, 13B, 30B, 70B, 175B, and so on, taking up tens or even hundreds of gigabytes.
On the other hand, LLM generative tasks require storing a large amount of KV cache, often exceeding the memory consumed by the weights, so as to enable sufficient parallelism and saturate GPU resources~\cite{yu2022orca, agrawal2024taming, zhong2024distserve, lee2024infinigen}.
To meet the requirement in terms of large memory, existing systems typically employ offloading or parallel approaches.

\begin{figure}[!t]
    \centering
    \includegraphics[width=0.95\linewidth]{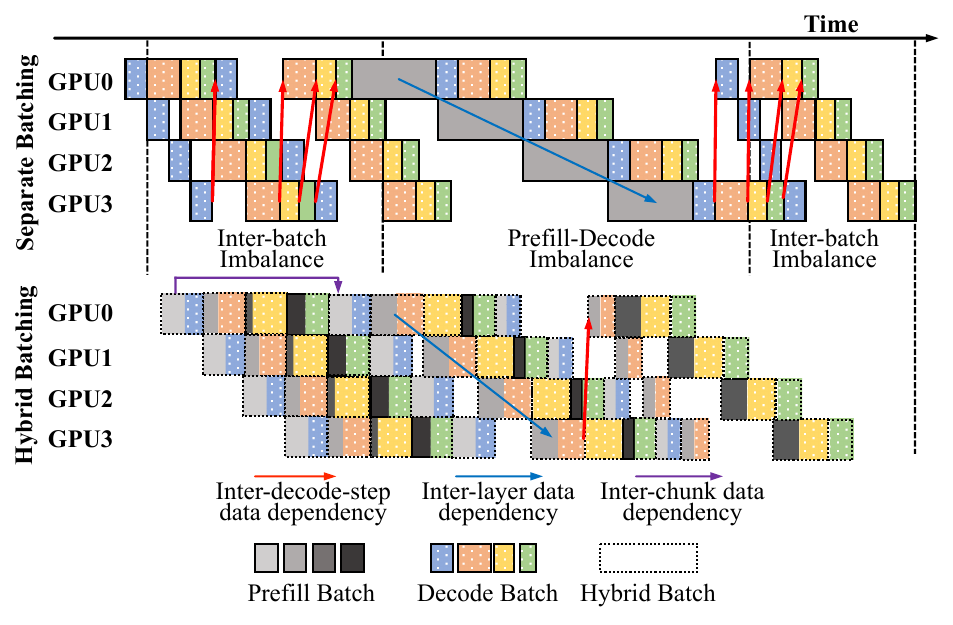}
    \caption{Bubbles in LLM pipeline parallel inference.}
    \label{fig:pipeline_bubble}
    \vspace{-10pt}
\end{figure}
Offloading approaches run an LLM with limited GPU memory by offloading data to the secondary storage and retrieving it when the data is accessed. 
By carefully optimizing the computation schedule and tensor placement, existing offloading approaches~\cite{sheng2023flexgen, 2022deepspeed} can hide data transfer between GPU memory and the secondary storage with computation, achieving considerable throughput on each GPU.
Unfortunately, offloading approaches inherently suffer from severe contention problems.
Specifically, under the commonly-used multi-GPU architecture, several GPUs share the only one channel linked with CPU, and consequently there is serious bandwidth contention on CPU’s root complexes when multiple GPUs offload data simultaneously.

Parallel approaches distribute a single LLM inference instance across multiple devices.
Tensor parallelism~\cite{shoeybi2019megatron, dudu2024liger} is widely used in existing LLM serving systems~\cite{nvidia2023fast, 2022deepspeed, vllm2024, tensorrt2024} due to its simplicity.
Through splitting each layer into symmetric fragments and distributing fragments, devices performs the similar computation, and they communicate with collective primitives, following the single program multiple data (SPMD) model.
However, due to the frequent and intensive synchronizations, considerable communication overhead makes it inefficient.

In comparison, pipeline parallelism~\cite{huang2019gpipe, dudu2024liger} splits a model by layers and only transfers activations between cut-off layers.
This low-communication feature makes it a promising approach for fully exploiting the computational capabilities of devices.
However, due to the imbalanced pipeline workloads and complex data dependencies, achieving compact scheduling is challenging, and bubbles always exist.
Figure~\ref{fig:pipeline_bubble} presents the running process of batching prefill and decode phases separately and scheduling them in an interleaved manner, as in most practical LLM inference systems~\cite{vllm2024, tensorrt2024,zheng2024sglang}.
The imbalanced workloads come from two aspects.
First, different batch sizes lead to different decode step workloads, resulting in the inter-batch imbalance.
Second, the prefill-decode imbalance exists as the execution of prefill usually takes much longer time than a decode step. 
Likewise, the data dependencies also come from two aspects.
On the one hand, inter-layer data dependency requires data to be processed in the current layer before entering to the next layer for processing.
On the other hand, inter-decode-step data dependency exists as the execution of generating the next token cannot start until the previous iteration completes.

\begin{figure}[!t]
    \centering
    \includegraphics[width=0.95\linewidth]{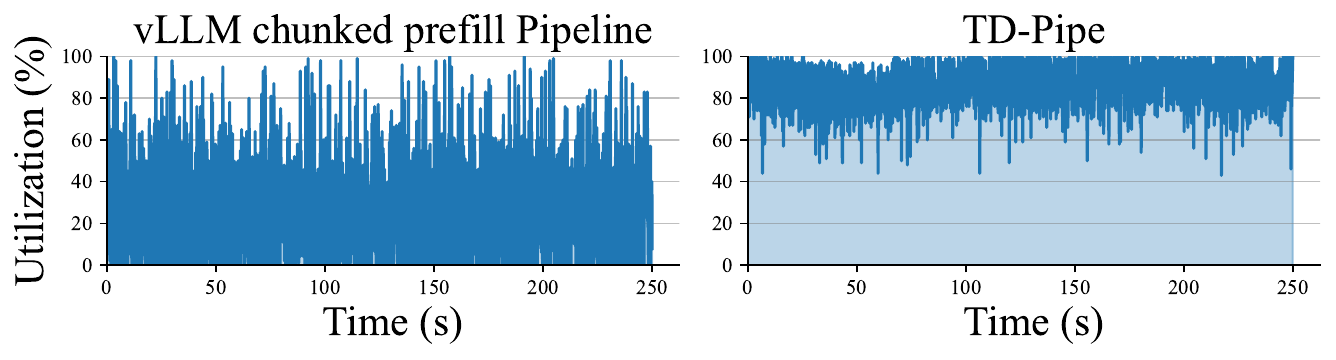}
    \caption{GPU Utilization in Pipeline Parallelism.}
    \label{fig:gpu_usage}
    \vspace{-10pt}
\end{figure}

An alternative approach to mitigate the pipeline bubble problem is the combination of the chunked-prefill approach~\cite{agrawal2024taming} and the hybrid batching~\cite{yu2022orca}. 
Chunked prefill is now commonly used in many LLM serving systems to balance TTFT and TPOT.
By splitting the prefill computation into chunks and mixing these chunks with decode, chunked prefill can achieve better inter-batch load balance.
However, 1) it depends on the prefill-to-decode ratio and struggles to handle requests with continuously varying lengths, 2) it introduces tighter data dependency as prefills are mixed with decodes and chunks of a single prefill introduce new data dependency, 3)it incurs repeated KV cache loading overhead.
The lower part of Figure~\ref{fig:pipeline_bubble} illustrates how its bubbles emerge even when starting from an initially balanced pipeline.
Over time, complex data dependencies and the continuous arrival and completion of requests exacerbate this issue.
To better illustrate its inefficiency, the left part of Figure~\ref{fig:gpu_usage} shows that it suffers from significant GPU underutilization.
Therefore, existing pipeline parallelism approaches always suffer from severe bubble problems.

To better leverage pipeline parallelism for high-throughput LLM inference, we propose TD-Pipe, an LLM inference system based on the temporally-disaggregated pipeline parallelism architecture. Its key idea is to disaggregate the prefill and decode phases in the temporal dimension, or rather, to reduce the prefill-decode switching frequency as much as possible.
Thus, load balance between pipeline stages can be effectively maintained in prefill and decode phases respectively and data dependency can be largely resolved.
As shown in right part of Figure~\ref{fig:gpu_usage}, TD-Pipe significantly improves GPU utilization.

We further identify key challenges in supporting this novel architecture and provides solutions.
It includes a hierarchy-controller system structure and three novel approaches.
The hierarchy-controller architecture enables asynchronous device coordination in a pipeline.
Since different devices perform consecutive layers of a model and LLM inference workloads are generally irregular in terms of input and output lengths, the device-to-device transfer has to be in a blocking style.
The hierarchy-controller decouples the scheduling from execution by introducing the distributed runtime (execution plane) and the centralized engine (control plane), enabling unblocked transmission.

Next, we present three approaches to further enhance the temporal disaggregation benefit.
First, to delay the timing of switching from prefill to decode and perform more prefills, we propose the AI-based greedy prefill approach.
To prevent it from exceeding the memory capacity, we leverage the output length prediction model.
As some requests will finish and the related KV cache is freed during decoding, we can perform prefills more aggressively with length prediction. 
Second, to reduce pipeline bubbles in the decode phase, we propose the inter-batch work stealing approach.
Since requests will finish randomly and result in imbalance, this approach dynamically balances the number of requests between batches.
Third, to determine the optimal timing of switching from decode back to prefill, we propose the spatial-temporal intensity comparison approach.
As decode progresses, some requests complete, leading to a gradual drop in computational intensity, this is referred to as spatial intensity.
Meanwhile, switching to the decode phase introduces pipeline bubbles, and the proportion of time the hardware is effectively utilized is defined as temporal intensity.
By comparing them, we identify the switching point.
In summary, we make the following contributions:
\begin{itemize}[leftmargin=*]
    \item We identify causes of bubble problem in LLM generative tasks when using the pipeline parallelism. 
    \item We design the novel temporally-disaggregated pipeline parallelism architecture targeting at throughput-oriented offline LLM inference, which disaggregates the prefill and decode phases in the temporal dimension.
    \item We identify potential issues with the exploitation of novel architecture and present TD-Pipe with three approaches.
    \item We evaluate TD-Pipe and show its efficiency.
\end{itemize}
\section{Background}
\subsection{LLM Generative Tasks}
\label{sec:llmdecoding}

\begin{figure}[!t]
    \centering    \includegraphics[width=0.75\linewidth]{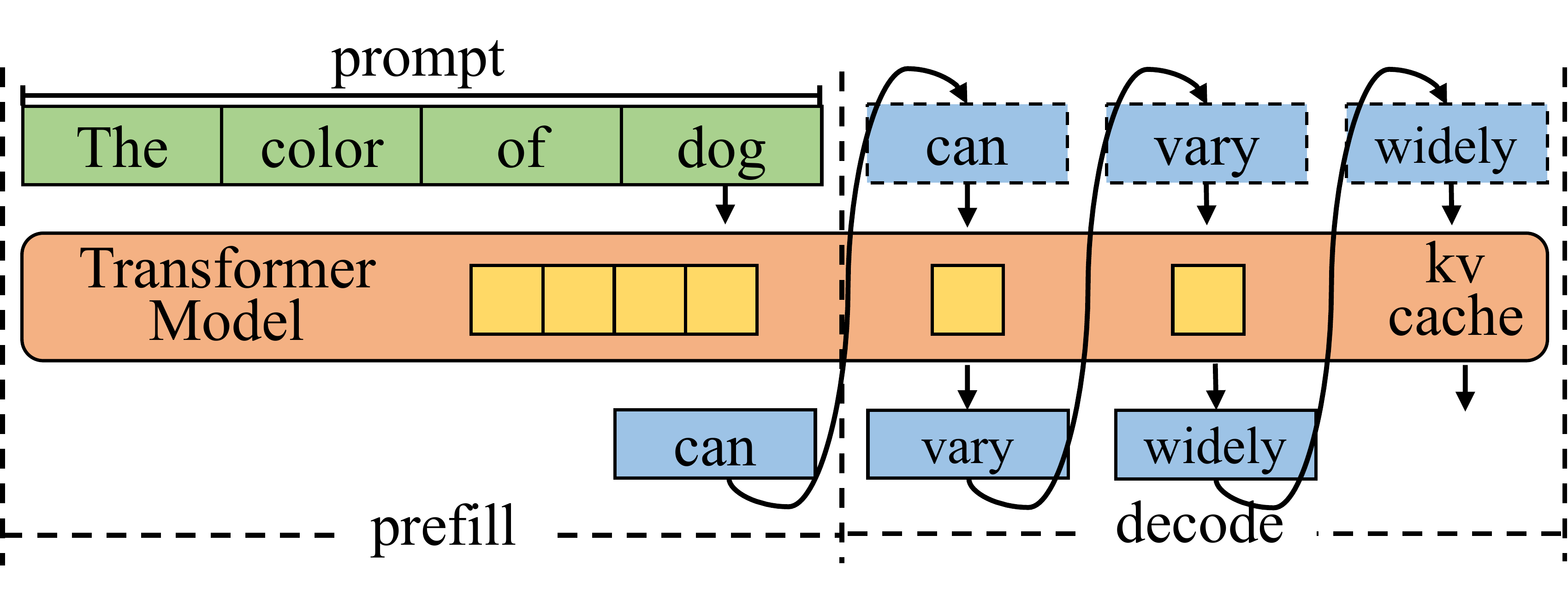}
    \caption{The generate process of LLM generative task.}
    \label{fig:gen-task}
    \vspace{-8pt}
\end{figure}
Modern LLM generative tasks follow the autoregressive decoding process and predict the next token given an input sequence.
During this, intermediate states, known as KV cache, are generated at each token position.
Through maintaining KV cache in memory, the modern LLM inference systems can eliminate most redundant computation in this iterative process.
Based on whether the KV cache has been generated, the inference process is divided into prefill and decode phases.
As illustrated in Figure~\ref{fig:gen-task}, the prefill phase deals with the newly-incoming prompt (in green) and initializes the KV cache (in gold), often comprising many tokens, and processes these tokens concurrently to generate a new token (in blue).
Next, benefiting from KV cache, each step of the decode phase only processes one new token from the previous step. LLM generative tasks have some key features:
\begin{itemize}[leftmargin=*]
    \item First, the prefill and decode have significant computational differences. For the prefill, it processes many tokens in parallel and easily becomes compute-bound. In contrast, the decode is more constrained by the memory bandwidth for it has a similar amount of I/O and only processes a single token for a single request. Thus, a very small batch size is sufficient for the prefill phase to saturate computational resources, while the decode phase requires a substantially larger batch size.
    \item Second, inherent uncertainty arises from the variable-length nature. Both input and output lengths vary, and the output length typically remains unknown until inference completes.
\end{itemize}

\subsection{Opportunities of Pipeline Parallelism for High-Throughput LLM Inference}

\subsubsection{High-Throughput LLM inference}

While online LLM serving has garnered significant attention, scenarios without strict latency constraints, such as the RLHF rollout stage~\cite{hu2024openrlhfeasytousescalablehighperformance} and data augmentation for recommendation systems~\cite{ding-etal-2024-data}, are also important for modern intelligent applications.
In such scenarios, pursuing high throughput becomes the top priority. 
\begin{figure}[t]
    \centering    \includegraphics[width=0.75\linewidth]{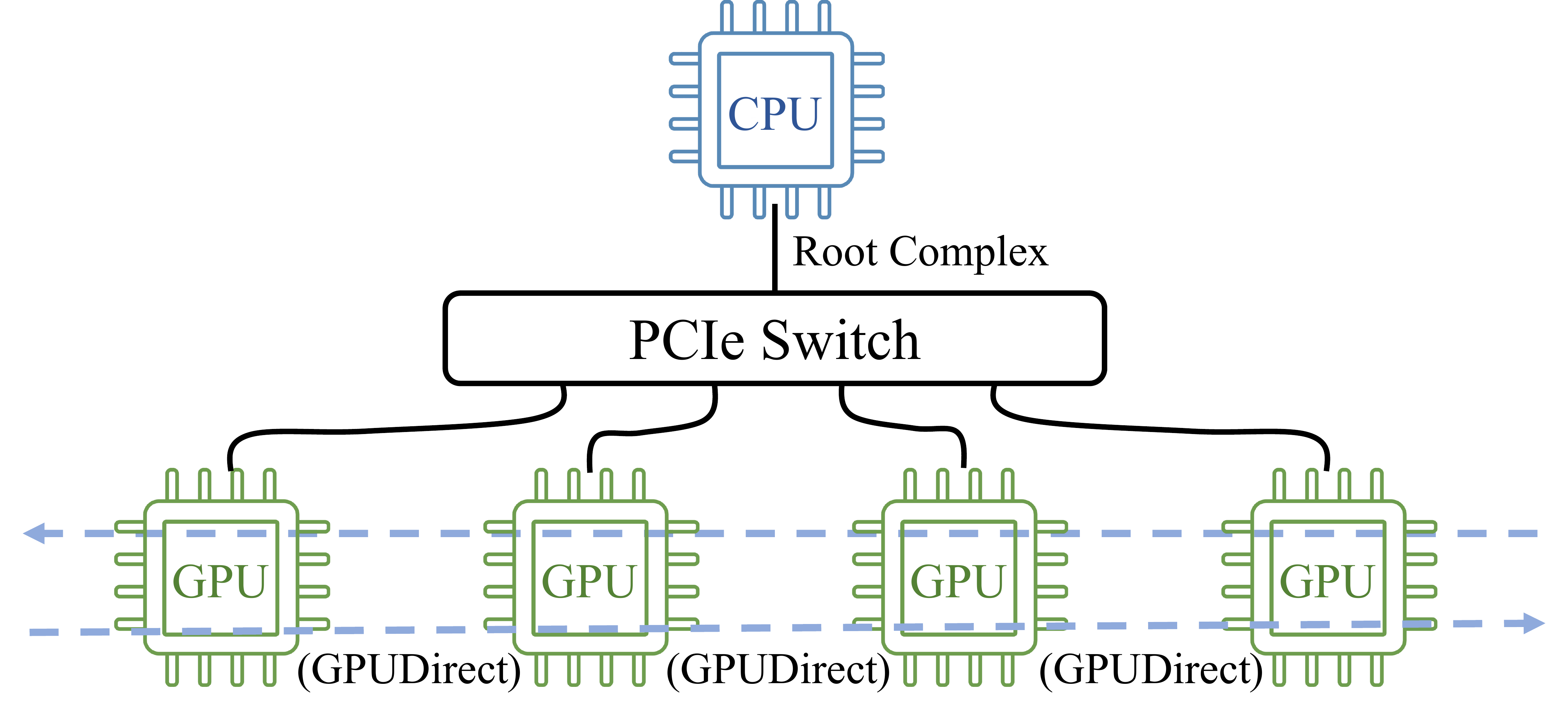}
    \caption{The multi-GPU architecture.}
    \label{fig:multigpu}
\end{figure}

First, achieving high-throughput LLM inference requires sufficient memory capacity.
During inference, both model weights and KV cache demand substantial memory.
Commonly-used LLMs have 7B, 13B, 30B, 70B, or even more weights, exceeding the memory capacity of most devices, especially these commodity devices such as NVIDIA A10 (24GB), 4090 (24GB), and L20 (48GB).
In addition to weights, hundreds of requests should coexist in the decode phase for saturating GPU resources and KV cache contributes significantly to memory usage. 
For example, the KV cache of a single token in the Llama-30B~\cite{touvron2023llama} occupies 1.52 MB, and 400 requests with an average length of 300 would require approximately 178 GB of memory. 

Second, achieving high-throughput LLM inference requires minimizing coordination overhead when utilizing multiple resources to expand memory capacity.
Figure~\ref{fig:multigpu} illustrates the popular multi-GPU architecture widely used in deploying LLMs, where multiple GPUs connect to a single CPU.
Under this architecture, there are typically two major approaches to improve memory capacity: offloading and parallelization.

\begin{figure}[!t]
    \centering    \includegraphics[width=0.95\linewidth]{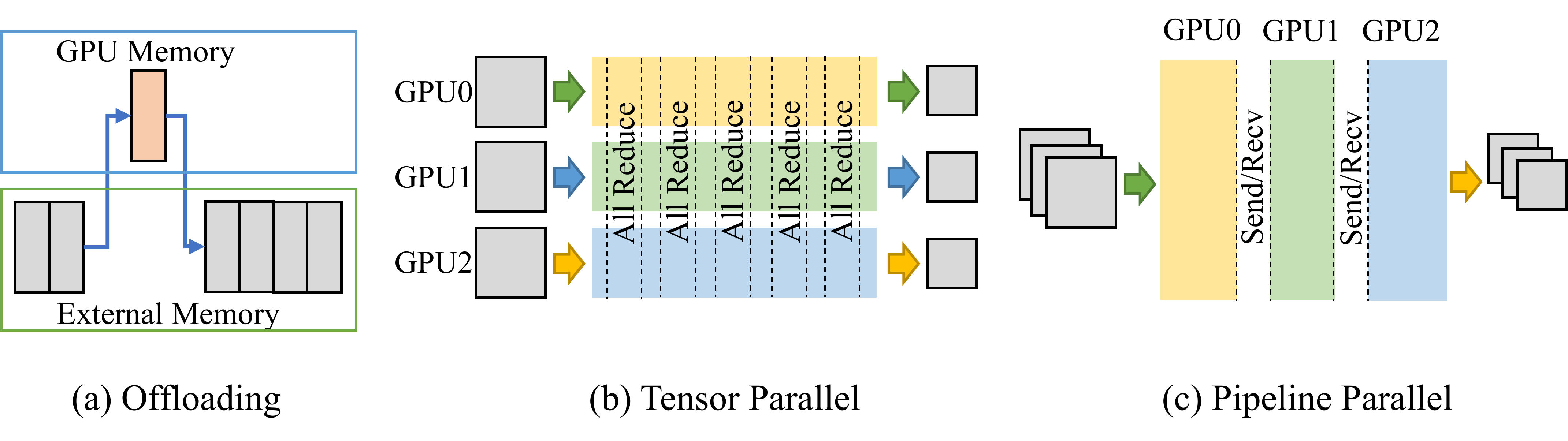}
    \caption{Offloading and parallel approaches.}
    \label{fig:moremem}
\end{figure}

\subsubsection{Coordination Overhead in Offloading Approach}

As shown in Figure~\ref{fig:moremem}(a), a single inference instance can acquire more memory capacity through offloading data to secondary storage.
Through iteratively exchanging data between host memory and GPU memory, this allows the decode phase to accommodate more requests for processing at a given moment.
However, the PCIe bandwidth can be the bottleneck for this approach even when the CPU-to-GPU ratio is 1:1.
It encounters more severe PCIe contention in a multi-GPU architecture when multiple GPUs simultaneously offload data to host memory.
In other words, the substantial coordination overhead between CPU and GPU makes this approach infeasible for high-throughput LLM inference.

\subsubsection{Coordination Overhead in Parallel Approaches}

\begin{figure}[!t]
    \centering    \includegraphics[width=0.95\linewidth]{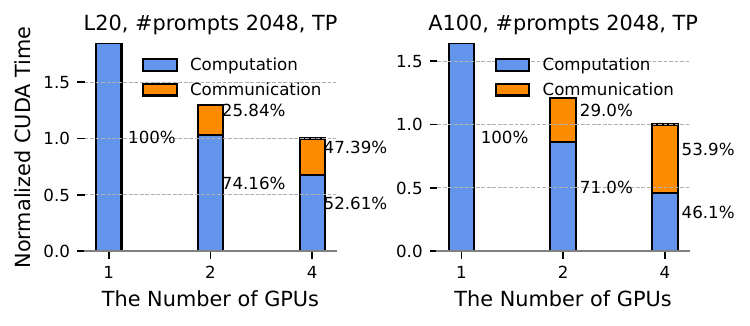}
    \caption{Execution time breakdown of the prefill phase using the tensor parallel approach in a multi-GPU architecture.}
    \label{fig:background_TD}
\end{figure}

Through distributing LLMs to multiple devices, an instance can acquire more memory capacity.
Figure~\ref{fig:moremem}(b) and (c) show the tensor parallel (TP) and pipeline parallel (PP) approaches.
TP shards each layer across multiple GPUs, with the model weights and KV cache being equally distributed across GPU workers.
It demands frequent communications that a single Transformer layer necessitates two rounds of all-reduce operations.
We deliver distributed inference case studies of Llama-30B~\cite{touvron2023llama} on two nodes without direct cables in FP16.
To demonstrate strong scalability, we reduce the number of layers in the model to fit into fewer devices and report the running results, shown in Figure~\ref{fig:background_TD}.
Notably, the model is composed of layers with the same structure, so reducing the number of layers does not affect its computational or communication characteristics.

The overall execution time on the L20 node is reduced by 1.84$\times$ when increasing the number of devices from 1 to 4.
Despite the reduction in total execution time, communication time accounts for 47.39\% of the total time on 4 devices.
On the A100 node, the overall execution time is only reduced by 1.64$\times$, with communication time consuming up to 53.9\% of the total time on 4 devices.
Thus, the tensor parallel approach heavily relies on interconnection capability and leaves computational resource idle for long period. 
Furthermore, certain high-performance GPUs enable direct communication through specialized cables, like NVIDIA's NVLINK, which significantly enhances their communication efficiency but also largely raises the node price.

Compared to TP, PP splits a model layer-wise, where each device is responsible for a subset of layers. 
It only requires a single point-to-point communication with a much smaller data volume every few layers.
Although the imbalanced pipeline workloads and complex data dependencies of LLM inference still prevent it from being the primary choice in the latency-sensitive scenarios,  the largely-reduced communication makes PP a promising approach to support throughput oriented LLM inference, especially on commodity hardware that high-performance connectivity like NVLINK is unavailable.

\subsection{Existing Pipeline Parallelism for LLM Inference}
\label{sec:batching}
Existing pipeline parallelism mainly differs in batching approaches.
To fully utilize modern GPUs, the batching technique is commonly adopted for processing deep learning workloads, where multiple samples are processed simultaneously to expose high parallelism and provide considerable hardware performance improvements.
There are mainly separate batching and hybrid batching for LLMs.

Separate batching schedules requests exclusively for the prefill phase or exclusively for the decode phase, without mixing them in the same batch. Given that these two phases exhibit distinct performance characteristics, i.e. compute-bound and memory-bound, the prefill phase saturates GPU computation even at a batch size of just one, while the decode phase requires a batch size in hundreds.
At each time, the inference instance will stay either in the prefill phase or in the decode phase.
By contrast, hybrid batching combines requests from both the prefill and decode phases into the same batch, forming hybrid batches.

When integrated with pipeline parallelism, these two batching approaches exhibit distinct advantages and disadvantages, yet both continue to suffer from severe pipeline bubble issues.
With separate batching, it is straightforward to manage batches consisting solely of prefill-phase requests, which can be launched continuously since there are no inter-batch data dependencies. Similarly, decode-only batches are convenient for achieving inter-batch load balance, as each decode step processes a single token with relatively uniform workload. However, mixing prefill and decode requests within the same batch introduces data dependencies and workload imbalance, which can lead to severe pipeline bubbles.
Moving to hybrid batching, it is typically integrated with the chunked prefill approach~\cite{agrawal2024taming}, which splits the prefill into chunks and enables to achieve better inter-batch load balance.
However, 1) as decode steps are added to all batches, it becomes essential to manage data dependencies across all batches, 2) the highly variable input and output lengths still lead to severe load imbalance, 3) Moreover, chunked prefill introduces repeated KV cache access as overhead.

\section{The Design of TD-Pipe}

\begin{figure*}[htbp]
    \centering    \includegraphics[width=0.95\linewidth]{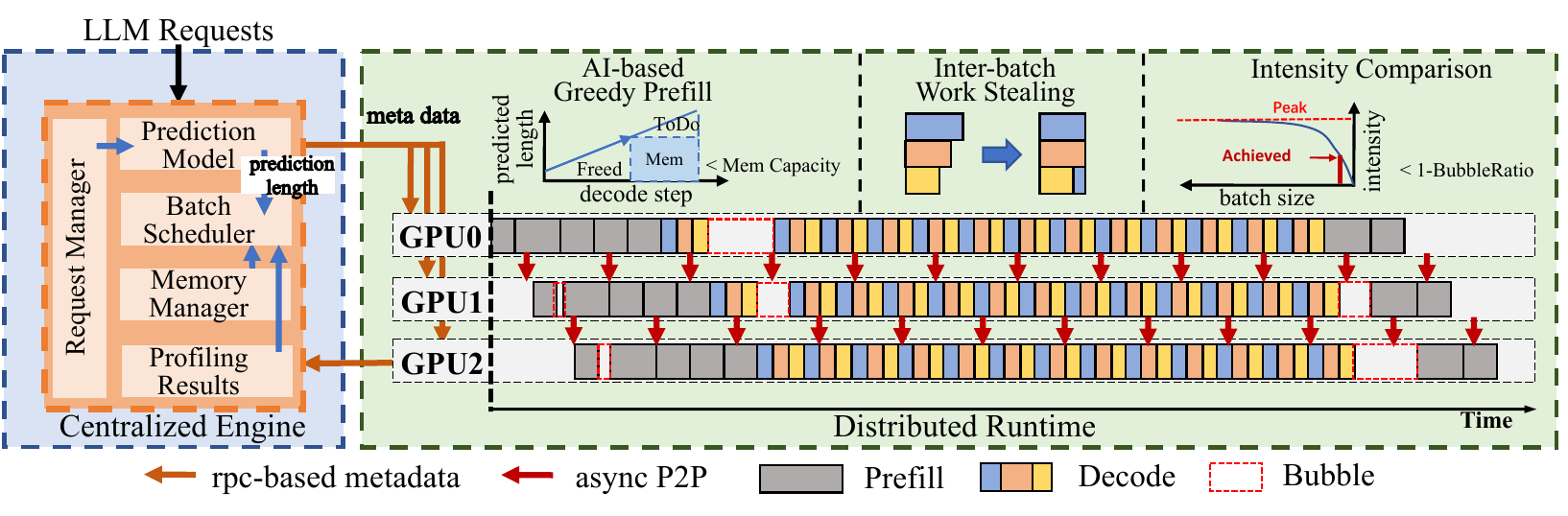}
    \caption{The system overview of TD-Pipe.}
    \label{fig:overview}
\end{figure*}

The key idea of TD-Pipe lies in the temporally-disaggregated pipeline parallelism architecture which disaggregates the prefill and decode phases in the temporal dimension.
Basically, TD-Pipe aims to achieve high-throughput offline LLM inference in the multi-GPU architecture, particularly on commodity nodes without GPU-direct interconnection cables.
Figure~\ref{fig:overview} demonstrates the overall design of TD-Pipe.
At first, TD-Pipe follows a hierarchy-controller structure, which abstracts the system into a centralized engine and a distributed runtime, responsible for controlling and execution respectively. 
Through decoupling the scheduling from execution, it can coordinate multiple devices in a pipeline asynchronously.

Next, TD-Pipe identifies three fundamental requirements in the temporally-disaggregated pipeline parallelism architecture and addresses them by three novel approaches.
First, the AI-based greedy prefill approach aggressively delays and determines the timing for switching from prefill to decode.
It performs more prefills within a fixed memory capacity by predicting output lengths and calculating the optimal point.
Second, the inter-batch work stealing approach reduces pipeline bubbles caused by random request completion in different batches and dynamically redistributes workloads from heavy to light batches.
Third, the spatial-temporal intensity comparison approach  determines the timing for switching from decode to prefill by comparing the performance cost between the reduced computational intensity and the pipeline bubbles caused by switching.
Finally, TD-Pipe achieves compact pipeline scheduling with very minimal bubbles in phase switching.

\subsection{Temporally-Disaggregated Pipeline}

As stated in Section \S \ref{sec:batching}, naively integrating separate batching into pipeline parallelism leads to severe bubble issues, primarily due to data dependencies and workload imbalance.
To address this, we adopt a temporal disaggregation strategy that separates the prefill and decode phases in the temporal dimension, thereby minimizing their mutual interference.
In this design, each inference instance processes only a single phase type over a long period of time, significantly reducing pipeline bubbles.
The prefill phase and the decode phase operate in a producer-consumer relationship, where the decode phase relies on the completion of the prefill phase.
Consequently, phase switching remains necessary.
To further improve this architecture, we aim to reduce the frequency of phase switching while maintaining high efficiency within each phase.

\subsection{Hierarchy-controller System Structure}

To make devices collaborate better, we design the hierarchy-controller system structure for TD-Pipe.
To handle the complex data dependencies and enable asynchronous data transmission simultaneously, as shown in Figure~\ref{fig:overview}, it decouples the scheduling and execution of pipeline tasks, abstracting the centralized engine as the control plane and the distributed runtime as the execution plane.
The centralized engine is responsible for batch scheduling based on memory capacity, request status, and profiling results, and then launching tasks to workers of distributed runtime.
The distributed runtime is responsible for detailed execution and each worker only knows what data and which device it should communicate.

\subsubsection{Centralized Engine}
The centralized engine manages and controls workers via remote procedure call (RPC).
It is primarily responsible for two distinct stages, i.e. runtime initialization and execution scheduling.
During initialization, it delegates sub-models to workers, initializes the relevant parts of the model and loads parameters into memory.
Execution scheduling deals with batching requests and sending control information to workers.
For the prefill phase, the engine packs prompts and directly sends them to workers until the switching condition is triggered.
For the decode phase, the engine packs decode steps as batches and is also responsible for deleting these completed requests at each iteration.

\subsubsection{Distributed Runtime}

As only the execution plane, the distributed runtime follows the idea of the single program multiple data (SPMD) model, that each worker only knows what data it should compute, what data it should communicate, and which device it should communicate to.
To manage the communication meta-information, such as world size and rank number, a global communication context is initialized.
In this way, after receiving information distributed by the centralized engine, a worker knows its position based on the global communication context, and it can perform the point-to-point communication between pipeline stages asynchronously.

\subsection{Approach-1: AI-based Greedy Prefill}

\begin{figure}[!t]
    \centering    \includegraphics[width=0.95\linewidth]{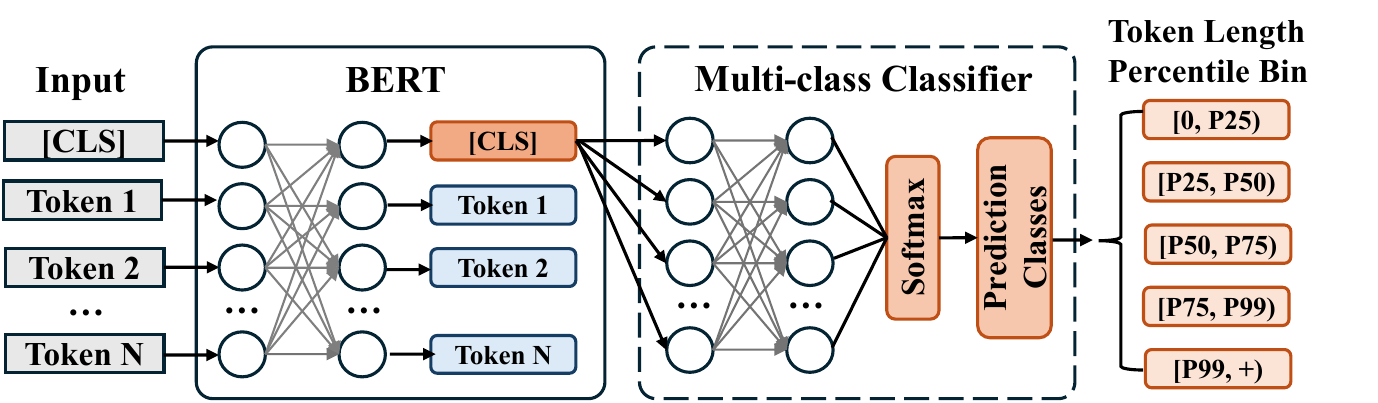}
    \caption{Prediction model for output length}
    \label{fig:proxy-model}
    \vspace{-8pt}
\end{figure}

\begin{algorithm}[!t]
\setlength{\textfloatsep}{0.5cm}
\setlength{\floatsep}{0.5cm}
\setlength{\intextsep}{-5em}
\KwData{\textbf{kvUsage}: Predicted kv cache usage map at future sampling points.  \textbf{kvCapacity}: System kv cache capacity. \textbf{futurePoints}:  Step points for future decision-making}
\SetKwFunction{CheckSwitch}{CheckSwitch}
\SetKwProg{Fn}{Function}{:}{}
\SetKwFunction{updateUsage}{UpdateUsage}
\SetKwFunction{schedule}{SchedulePrefill}
\Fn{\CheckSwitch{kvUsage,  kvCapacity}}{
    maxUsage $\gets$ 0 \;
    \ForEach{usage $\in$ kvUsage}{
        \If{usage > maxUsage}{
            maxUsage $\gets$ usage \;
        }
    }
    \If{maxUsage > kvCapacity}{
        Switch to decode\;
    }
    \Else{
        Remain in prefill\;
    }
}
\Fn{\updateUsage{prefillRequest, kvUsage}}{
    inputLen $\gets$ getInputLen(prefillRequest)\;
    predictLen $\gets$ getPredictLen(prefillRequest)\;
    \ForEach{futurePoint $\in$ futurePoints}{
        \If{futurePoint $\leq$ predictLen}{
            kvUsage[futurePoint] += (inputLen + futurePoint)\;
        }
    }
}

\Fn{\schedule{kvUsage,  kvCapacity}}{
    batch = getPrefillBatch().Launch()\;
    \ForEach{request $\in$ batch}{
        \updateUsage{request, kvUsage}
    }
    \CheckSwitch(kvUsage,  kvCapacity)
}
\caption{Prefill-to-Decode Switch Algorithm}
\label{algo:memusage}
\end{algorithm}

After performing the prefill phase, TD-Pipe requires to determine the timing for switching from prefill to decode.
To reduce the switching frequency between the prefill and decode phase, TD-Pipe tends to execute more prefills before switching to the decode phase.
However, the decode phase also generates KV cache and consumes memory, blindly performing too many prefills will lead to frequent re-computation or offloading in the decode phase, which results in a significant performance drop.
Unfortunately, output lengths are unknown until requests complete and the peak memory usage is also unknown.

To determine how many prefills should be launched without largely exceeding the memory capacity during the decode phase, we propose the AI-based greedy prefill approach.
If we could know input and output lengths of all requests, we could calculate the total memory usage and determine the switching timing.
Furthermore, during the decode phase where all requests progress together, some requests generate new KV cache while others complete, freeing the associated KV cache.
By simulating the memory usage throughout the entire process, we can perform prefills more aggressively. 

To accurately predict the output length for each request, we follow the prediction model introduced in µ-Serve ~\cite{qiu2024muserve}, as illustrated in Figure~\ref{fig:proxy-model}. This model employs a BERT-based multi-class classifier to estimate the output token length for each user request. Specifically, it utilizes the hidden state of the [CLS] token from the final layer of BERT as an aggregate representation of the input, which is then passed through a two-layer feedforward neural network to predict the output length category. These categories are defined as ranges, starting from [P0, P25) to [P99, +), representing the percentage intervals derived from historical inference data. We assume that the inference inputs within a given scenario exhibit strong similarities, allowing us to train the model based on historical patterns. For each request, the predicted output length is assigned based on the average value of the category from the training set. 
Notably, the execution overhead of the prediction model is extremely low and can be neglected during the entire inference process.

With output length predictions, we design a dynamic programming (DP) algorithm, as in Algorithm~\ref{algo:memusage}, to efficiently simulate the memory usage for the subsequent decoding steps after performing new prefills. This simulation is performed after each prefill completes, without introducing significant overhead.
To simplify the computational complexity, we designate certain decode steps in the future for decision-making, referred to as \textit{futurePoints}.
For example, we only check whether the memory usage in the 32nd, 64th, 96th, ..., 992nd, 1024th decode steps exceeds the memory capacity.
After newly launching a batch, the algorithm updates the KV cache usage at these decode steps (\textit{futurePoints}) by checking whether the predicted output lengths exceeds certain decode steps.
With the future KV cache usage, we can determine the timing for switching from prefill to decode by checking whether there exists a \textit{futurePoint} that the memory usage is larger than the memory capacity.

\subsection{Approach-2: Inter-batch Work Stealing}

\begin{figure}[!t]
    \centering    \includegraphics[width=0.95\linewidth]{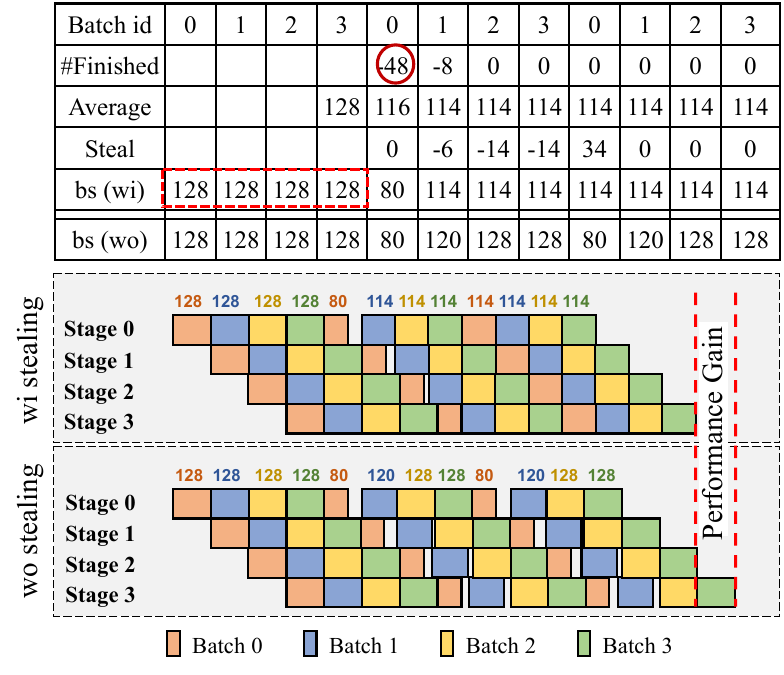}
    \caption{The inter-batch work stealing approach. \textbf{bs(wi)} represents the batch size with stealing and \textbf{bs(wo)} represents the batch size without stealing.}
    \label{fig:app-2}
\end{figure}

As the decode phase progresses, requests in different batches complete randomly and the workloads of different batches constantly change, resulting in imbalanced workloads.
Because of data dependencies between decode steps, it can result in pipeline bubbles.
The inter-batch work stealing approach is responsible for ensuring that workloads are evenly distributed across batches throughout the entire decode phase.

Initially, we need to determine when two decode batches can be considered to have equivalent workloads.
The execution time of a single decode step is determined by two factors: batch size and the lengths of KV cache.
The decode phase typically operates on large batch sizes, often in the hundreds, and KV cache lengths also vary significantly across requests.
This huge space makes it hard to accurately compare the execution time between two batches.
To simplify this, we consider batch size as the sole metric for indicating load balance between batches.
This is because linear operations generally dominate execution time in Transformer models across most scenarios and its execution time only relate to batch size.
Moreover, large batch sizes help smooth out execution time fluctuations caused by variation in sequence lengths.

At the beginning, we divide the requests into batches equal to the number of GPUs, with each batch containing the same number of requests.
As the decode phase progresses, requests in different batches complete randomly and the workloads of different batches constantly change, resulting in imbalanced workloads.
To make them balanced, the inter-batch work stealing approach requires to dynamically migrate workloads between batches.
Notably, the batch scheduler can observe the actual batch size of at most one batch at a moment, while other batches remain executing and some requests potentially complete.
Thus, this approach employs the idea of sliding window to balance inter-batch workloads on-the-fly.

When a batch completes and returns, the central scheduler firstly examines the batch and removes the finished requests.
Then, an average batch size is acquired by deducting the number of finished requests from the total number of requests maintained in the sliding window and calculating the average.
The sliding window, with length equal to the number of GPUs, tracks recent batch sizes.
If the current batch contains more requests than the average, the excess requests are temporarily withheld and the batch is submitted for further processing.
And in subsequent iterations, the scheduler will supplement other batches with withheld requests. 
Through this iterative process, the work stealing approach gradually redistributes workloads, effectively mitigating load imbalance. 

Figure~\ref{fig:app-2} demonstrates a slightly extreme example of how this approach gradually achieves load balance in a 4-stage pipeline.  
Initially, 512 requests are evenly divided into four batches of 128. 
After the first iteration, batch 0 completes 48 requests, leaving 80.
As shown in Figure~\ref{fig:app-2}, by applying a sliding window to the last four batches and subtracting the number of completed requests, we obtain an average batch size of 116.
Since the remaining 80 requests in batch 0 are below the current average of 116 requests per batch, all 80 are resubmitted.
8 requests in batch 1 completes and the current average batch size becomes 114. Since the number of requests in batch 1 exceeds 114, the scheduler steals 6 requests from batch 1 and submits only 114. For batch 2 and 3, no requests complete in the first round, and the scheduler similarly steals requests as needed to keep each batch close to the average. In the next iteration, these stolen requests are redistributed to batch 0, thereby achieving balanced workloads across all batches. 
In the lower part of Figure~\ref{fig:app-2}, the imbalanced workloads quickly trend towards balance, and the reduced pipeline bubble translates into performance gains.

\subsection{Approach-3: Spatial-Temporal Intensity Comparison}

\begin{figure}[!t]
    \centering    \includegraphics[width=0.95\linewidth]{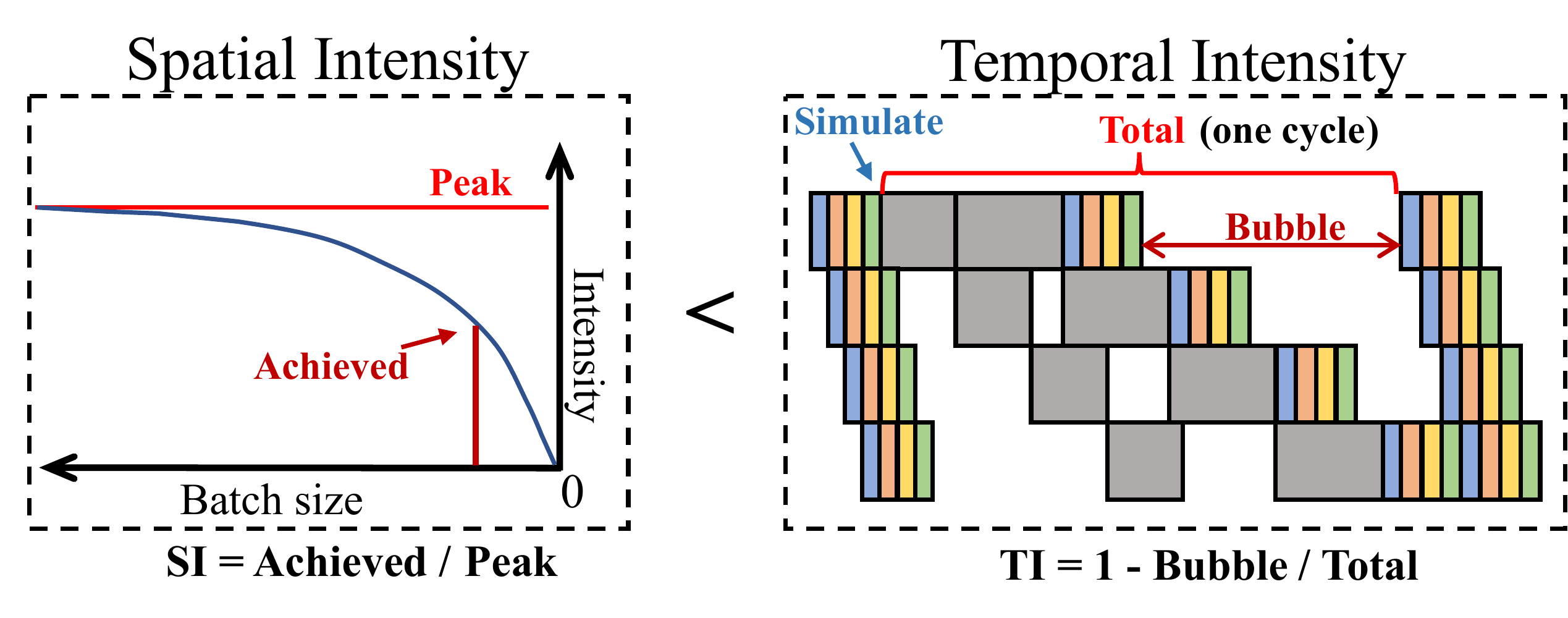}
    \caption{Spatial-temporal intensity comparison approach.}
    \label{fig:app-3}
\end{figure}

In the temporally-disaggregated pipeline architecture, the decode phase acts as the consumer. 
As the decode phase progresses, some requests complete, causing the batch size to decrease and leading to a drop in computational intensity.
If the computational intensity continues to drop, the decode phase would operate in low efficiency. 
However, if we switch to the prefill phase as soon as decode efficiency drops, frequent switching could create bubbles and harm performance.
To handle this dilemma, the spatial-temporal intensity comparison approach aims to determine the optimal timing for switching from decode to prefill.

As illustrated in Figure~\ref{fig:app-3}, the spatial-temporal intensity comparison approach evaluates and compares the computational efficiency of continuing with the decode phase versus switching to the prefill phase. To quantify these two efficiencies, we define spatial intensity for remaining in the decode phase and temporal intensity for switching to the prefill phase.

Spatial intensity is defined as the ratio of the current decode performance to the peak achievable performance.  In detail, we profile the execution time and calculate the average execution time per request using a sufficiently large batch size.
We refer to the reciprocal of the average execution time per request under high computational intensity as the \textit{Peak}. As shown on the left side of Figure~\ref{fig:app-3}, for each batch size, we similarly profile the reciprocal of the average execution time per request, which we denote as \textit{Achieved}. Spatial intensity is then  denoted by:
\vspace{-2pt}
\begin{equation}
\text{Spatial Intensity} = \frac{\textit{Achieved}}{\textit{Peak}}
\end{equation}
Then, temporal intensity is defined as 1 minus the bubble ratio.
The bubble ratio is the proportion of time lost to \textbf{bubble} if a switch occurs now, relative to the duration of the next prefill phase.
As shown in the right side of Figure~\ref{fig:app-3}, for each moment in the decode phase, we calculate the potential \textbf{bubble} and simulate the \textbf{total} time in the next cycle.
For the \textbf{bubble}, it is the difference between the longest prefill and the current decode.
For the \textbf{total} time, it is the accumulation of the pending prefills, each batch's one decode step (can even be ignored), and the bubble.
Since the pending prefills are determined, we can easily calculate their execution time.
Thus, temporal intensity is then denoted by:
\vspace{-2pt}
\begin{equation}
\text{Temporal Intensity} = 1 - \frac{\textit{bubble}}{\textit{total}}
\end{equation}

Based on these two intensities, we design the decision strategy as follows: when the spatial intensity is less than the temporal intensity, the system switches to the prefill phase. Otherwise, it continues with the decode phase.

\section{Evaluation}
\label{sec:evaluation}

We implement TD-Pipe on top of vLLM 0.5.3~\cite{vllm2024}.
We evaluate the effectiveness of TD-Pipe by comparing it with SOTA approaches and report the KV cache memory usage fluctuation during running.
Additionally, we conduct an ablation study to assess the impact of our proposed approaches.

\subsection{Experimental Setup}
\label{Setup}
\begin{table}[!t]
\caption{GPU Configurations}
\label{tab:gpus}
\centering
\resizebox{0.95\linewidth}{!}{
\begin{tabular}{c c c c c}
\hline
\textbf{  Device  }&  \textbf{   FP16 Tensor Core   } & \textbf{   Bandwidth   } & \textbf{   Memory   } & \textbf{   AllReduce   }\\ 
\hline \hline
L20 &  119.5 TFLOPS & 864 GB/s   &   48  GB      &   14.65 GB/s\\
A100 &  312 TFLOPS & 1,935 GB/s  &   80  GB      &   14.82 GB/s\\
\hline
\end{tabular}
}
\end{table}

\noindent\textbf{Node testbed.} 
Two multi-GPU nodes as in Table~\ref{tab:gpus} are employed.
One has 4 NVIDIA L20 GPUs (48GB) and another one has 4 NVIDIA A100 GPUs (80GB).
These two nodes all communicate through PCIe switch with a peak all-reduce bandwidth at 14.65 GB/s and 14.82 GB/s respectively.
Specifically, we use Pytorch$\sim$v2.3.1, vLLM$\sim$v0.5.3, CUDA$\sim$12.4, and NCCL$\sim$2.20.5.

\begin{table}[!t]
\caption{Model Specifications.}
\label{tab:models}
\resizebox{0.45\textwidth}{!}{
\begin{tabular}{ccccccc}
\hline
\textbf{Name}                        & \textbf{Parameters} & \textbf{Layers} & \textbf{Heads} & \textbf{Hidden Size} &\textbf{ Prec.} \\ \hline\hline
\multicolumn{1}{c}{Llama2-13B-chat} & 26GB & 40     & 40    & 5,120  & FP16       \\
\multicolumn{1}{c}{Qwen2.5-32B-Instruct} & 64GB & 64     & 40    & 5,120  & BF16     \\
Llama2-70B-chat                    & 140GB & 80     & 64    & 8,192  & FP16  \\ \hline
\end{tabular}
}
\end{table}
\noindent\textbf{Model setup.}
We choose Llama2~\cite{touvron2023llama2} and Qwen~\cite{qwen2025qwen25technicalreport} in different model sizes.
As detailed in Table~\ref{tab:models}, they are Llama2-13B-chat, Qwen2.5-32B-Instruct, and Llama2-70B-chat. For simplicity, we refer to them as 13B, 32B, and 70B in the following sections.
Notably, the 32B and 70B models use the group query attention (GQA) technique, which results in a smaller KV cache capacity for the same token count.

\begin{figure*}[htbp]
    \centering   
    \includegraphics[width=0.9\linewidth]{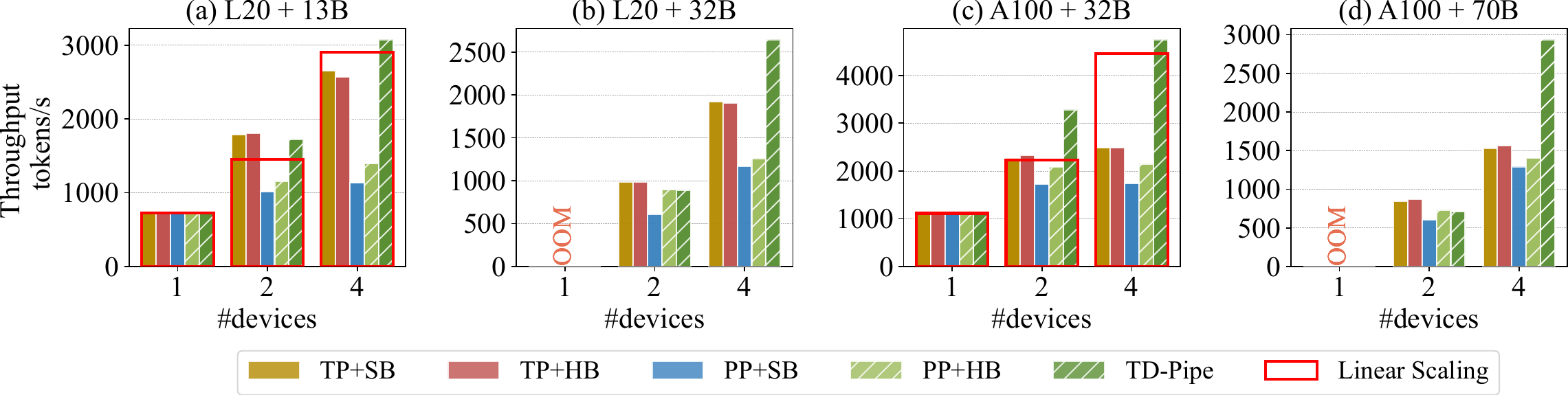}
    \vspace{-2pt}
    \caption{Overall performance.}
    \label{fig:overall-perf}
    \vspace{-10pt}
\end{figure*}
\noindent\textbf{Baseline Methods.}
We compare TD-Pipe with some recently released approaches implemented in vLLM~\cite{vllm2024}:
\begin{itemize}[leftmargin=*]
    \item \textbf{TP+SB}: The tensor parallel approach is integrated with separate batching. This method requires two All-reduce synchronizations in each transformer layer, and serves as the default approach in vLLM.
    \item \textbf{TP+HB}: The tensor parallel approach is integrated with hybrid batching, along with the chunked-prefill approach. This method also requires two All-reduce synchronizations in each transformer layer.
    \item \textbf{PP+SB}: The pipeline parallel approach is integrated with separate batching. This method requires limited communications but suffers from a severe bubble problem.
    \item \textbf{PP+HB}: The pipeline parallel approach is integrated with hybrid batching, along with the chunked-prefill approach. This method requires limited communications but still experiences the bubble problem.
\end{itemize}
Also, we employ the re-computation strategy where the KV cache of recently arrived requests will be freed once memory capacity is saturated, to deal with memory overflow issues.

\noindent \textbf{Dataset and Metrics.}
The popular conversation dataset, shareGPT V3~\cite{hfshareGPT}, is adopted in the evaluation.
The dataset contains around 53,000 samples, with each sample includes indefinite number of conversation rounds.
Following the benchmarking method used in vLLM, we filter input sentences with a length of less than 1024 tokens.
These input sentences are fed into the target models to generate new output sentences, constructing 86,612 input and output pairs.
For output length prediction training in the AI-based greedy prefill approach, 60\% of the samples are used as the training dataset, 20\% of the samples as the validation dataset, and all the other samples are used as the test dataset.
For performance evaluation, we randomly sample 5,000 input sentences as requests and use throughput (tokens/s) as the metric. 
Each execution of 5,000 requests takes around half an hour.

\subsection{Overall Performance Evaluation}

We compare TD-Pipe against baseline methods and present their overall throughput results for processing 5,000 input sequences.
Taking the ratio between memory capacity and model size into consideration, we select four node-model combinations: L20 + 13B, L20 + 32B, A100 + 32B, and A100 + 70B, and scale them from 1 to 4 GPUs.
We train output length predictor for these models respectively and report details in Section~\ref{sec:appeva1}.

Figure~\ref{fig:overall-perf} presents the throughput results, with the red box indicating the linear speed up.
In almost all cases, TD-Pipe presents advantages over other approaches, especially when the device number is 4.
For 4-device cases in all node-model configurations, TD-Pipe achieves up to 1.91$\times$ more throughput over TP+SB; 1.90$\times$ more throughput over TP+HB; 2.73$\times$ more throughput over PP+SB; and 2.21$\times$ more throughput over PP+HB.
Comparing PP+SB and PP+HB, it shows that the combination of hybrid batching and chunked-prefill approach can indeed optimize the pipeline parallelism by improving the inter-batch load balance. In contrast, TP+SB and TP+HB show fewer differences.

TD-Pipe demonstrates super-linear speedup as the number of GPUs increases from 1 to 4, largely outperforming other pipeline parallel approaches.
For example, when the GPU number scales from 2 to 4, the throughput of L20 + 32B grows 2.97$\times$.
When scaling from 2 to 4 GPUs, both the computing capability and the memory capacity are doubled.
On the one hand, the increased computing capability contributes to the speedup.
On the other hand, the increased memory capacity can be used for accommodating more KV cache and the computational intensity during the decode phase is largely improved.
Under the dual effects, TD-Pipe's throughput achieves super-linear improvement.
In comparison, the other two pipeline parallel approaches, PP+SB and PP+HB, show less improvement. This is because they are less impacted by memory capacity, and longer pipeline stages on more devices exacerbate their bubble problems.

Although TD-Pipe outperforms TP+SB and TP+HB in most cases, they can be constrained by various factors, i.e. memory capacity or coordination overhead.
Comparing L20 + 32B and A100 + 32B, we can first observe that A100 + 32B has significantly higher throughput than that of L20 + 32B, which aligns with the hardware specifications.
Additionally, we can also notice that the throughput ratio between TD-Pipe and TP methods varies between the two setups.
When performing with 4 GPUs, TD-Pipe achieves a performance improvement of 1.37$\times$ faster compared to TP+SB on the L20 node, while TD-Pipe is 1.90$\times$ faster than TP+SB on the A100 node.
This is related to many aspects, including the memory bandwidth, computing capability, memory capacity, and interconnection ability.
On the one hand, as shown in Table~\ref{tab:gpus}, the A100 is much more powerful than L20 in terms of both memory bandwidth and computing capability, while the interconnection ability is close.
Thus, TP+SB on the A100 node is more constrained by the interconnection, where TD-Pipe is much less reliant.
On the other hand, since all requests can be batched together 
using TP and requests are divided into multiple batches using PP, TP tends to achieve higher computational intensity than PP, while PP is more constrained by the memory capacity.
As a result, PP benefits more from the fact the A100 node's greater memory capacity.
Notably, since we record the overall throughput from the start of the first prefill to the finish of all decode batches, the execution tail of PP approach is less efficient compared to TP approach for its batch size is smaller. Nevertheless, our PP approach is still more effective than TP approach.

\subsection{KV Cache Memory Usage}

\begin{figure}[htbp]
    \centering    \includegraphics[width=0.9\linewidth]{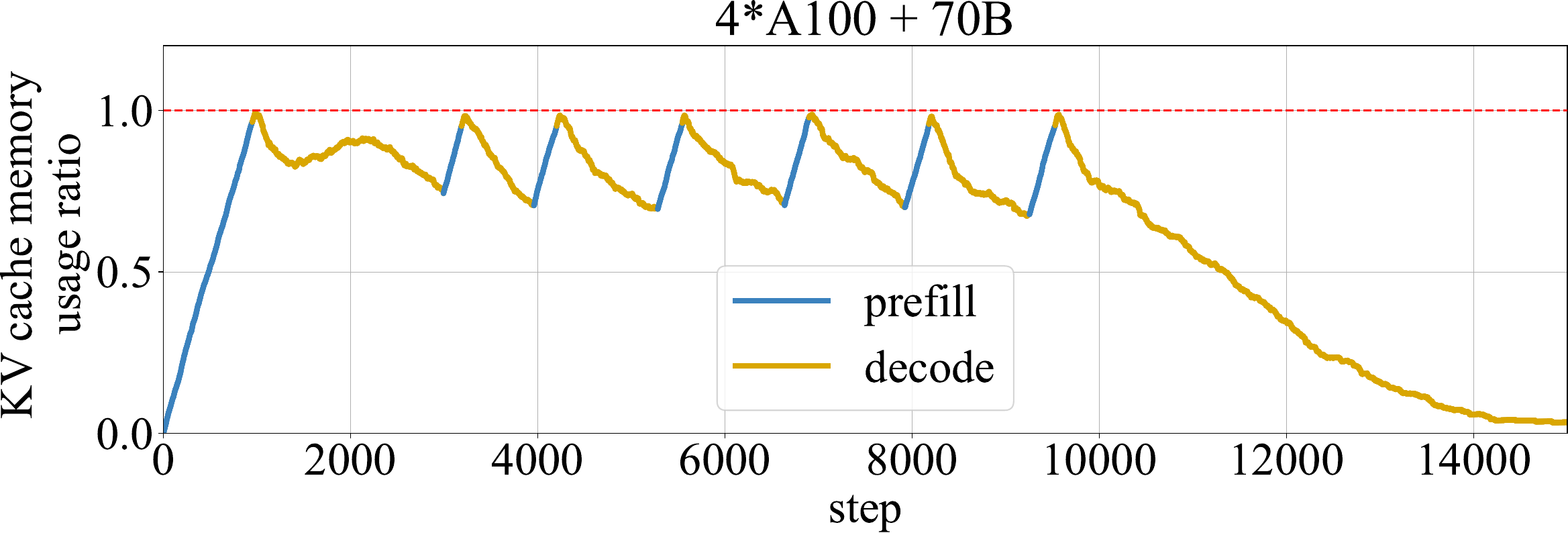}
    \vspace{-10pt}
    \caption{The KV cache memory usage fluctuation.}
    \label{fig:memory_usage}
    \vspace{-5pt}
\end{figure}

Figure~\ref{fig:memory_usage} illustrates the ratio of actual KV cache memory usage to the total allocated KV cache memory throughout the inference process, as the prefill and decode phases alternate over time.
Initially, KV cache memory usage increases continuously until the GPU memory approaches saturation, after which the system alternates between prefill and decode phases until all requests are completed.
For the prefill phase (in blue), the memory usage keep growing.
For the decode phase(in yellow), its memory usage always grows, fully occupied, and then begins to decline.
This is because new KV cache will be generated and some requests complete in the decode phase.
It is noted that the memory usage is only shortly occupied and it can illustrate the AI-based greedy prefill is effective.

\subsection{Ablation Study}
This section illustrates how each approach impacts TD-Pipe.

\subsubsection{Approach-1: AI-based Greedy Prefill}
\label{sec:appeva1}

\begin{figure}[htbp]
    \centering    \includegraphics[width=0.9\linewidth]{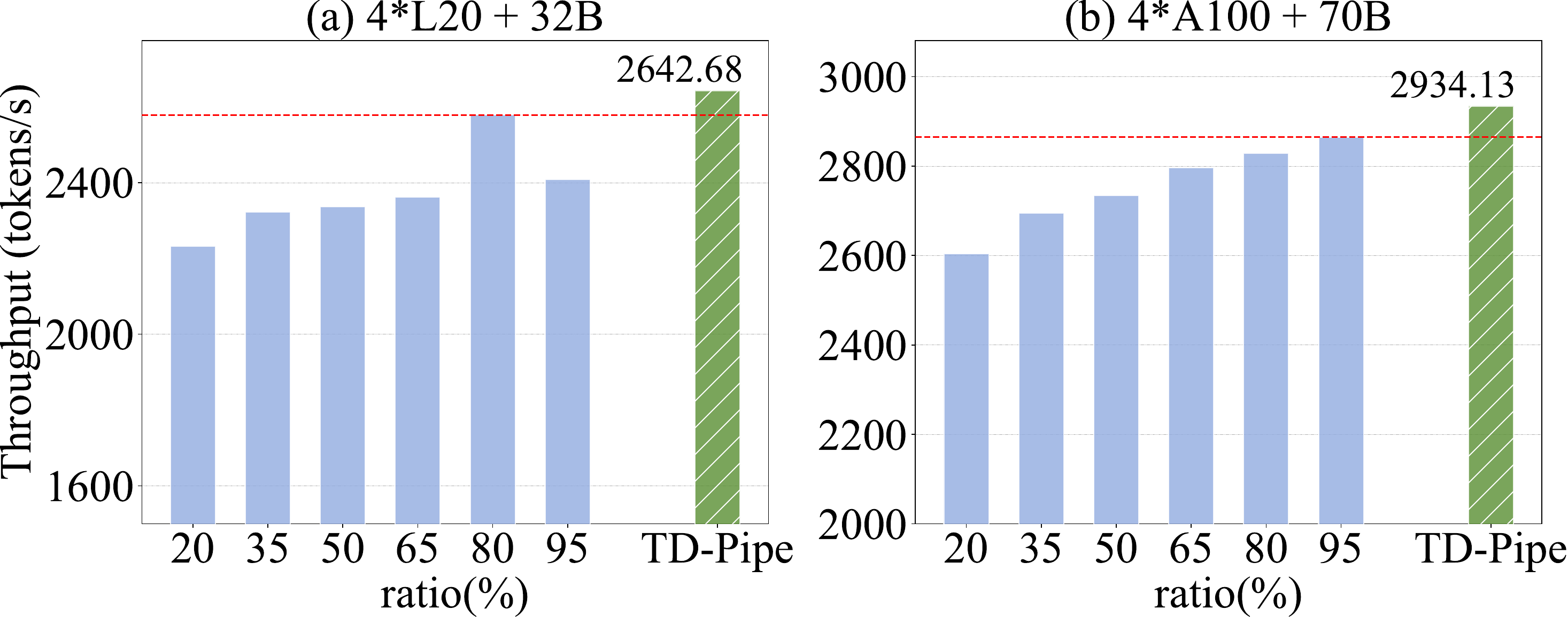}
    \vspace{-10pt}
    \caption{Ablation study for prefill-to-decode switching.}
    \label{fig:prefill-to-decode}
\end{figure}
To demonstrate how effective the AI-based greedy prefill can determine the timing for switching from prefill to decode, we design a hyperparameter called the KV cache occupancy ratio to replace the AI-based greedy prefill approach.
For example, when we set this ratio as 50\%, the prefill will switch to decode once 50\% KV cache blocks are occupied.
Figure~\ref{fig:prefill-to-decode} shows the throughput results, demonstrating that our approach outperforms all other manually selected switching points.
For the hyperparameter approach, it requires human experience and numerous experiments.
This result illustrates that our AI-based greedy prefill approach can determine a fairly good switching point, and it can dynamically fit the constantly changing requests and diverse hardware configurations.

\begin{figure}[htbp]
    \centering    \includegraphics[width=0.9\linewidth]{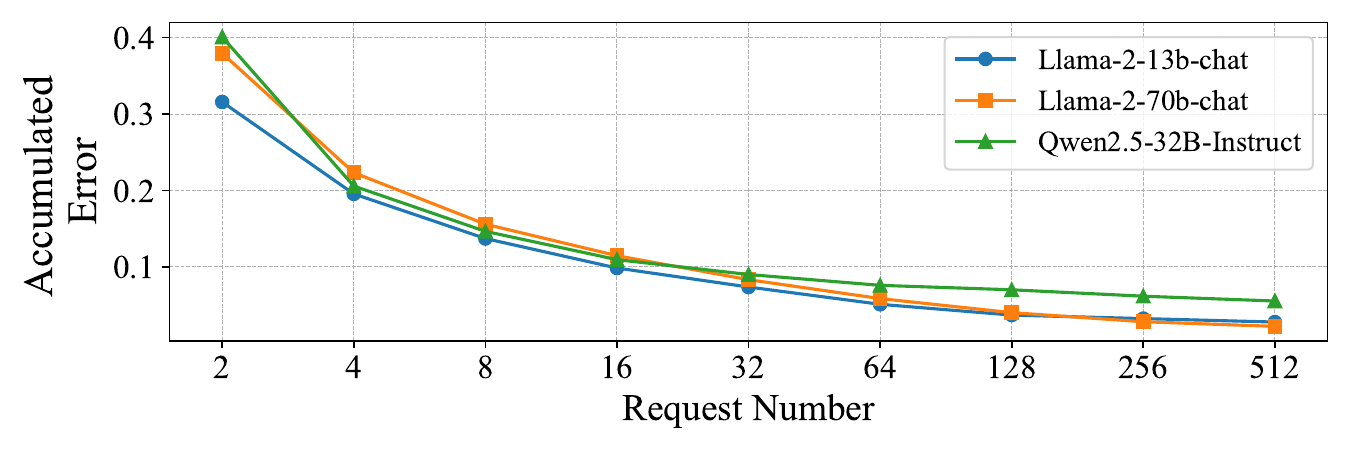}
    \vspace{-10pt}
    \caption{The accumulated error of varied request number.}
    \label{fig:error}
\end{figure}

Additionally, we report the prediction accuracy of the output length predictor.
The predictor is trained and deployed for each LLM model in FP16 precision. 
When predicting the length range of a single request, their accuracies are 0.5214, 0.5805, and 0.5234 for 13B, 32B, and 70B models, outperforming random guessing. 
Notably, single request's prediction accuracy can partially reflect its effectiveness in our scenario, while the accuracy of accumulated length prediction is more relevant.
Since some predictions may be overestimated while others are underestimated, the accumulated prediction accuracy can be smoothed out and better reflects the total KV cache size.
Therefore, as shown in Figure~\ref{fig:error}, we group samples in test set with different request numbers and evaluate the accumulated prediction error of each group.
The accumulated error is acquired by accumulating and averaging the relative difference between the predicted total length and the actual total length in all groups.
When the number of requests is large, the accumulated error becomes sufficiently small.
For example, when predicting the total output lengths of 256 requests, there are only 3.25\%, 6.18\%, and 2.84\%  errors for 13B, 32B, and 70B models.

Furthermore, we evaluate the overhead introduced by the predictor. For 5,000 requests, the predictor takes 1,418.861 ms on the L20 node and 833.695 ms on the A100 node.
In the experiments described in Section~\ref{Setup}, the shortest total processing times observed are 929 seconds on the L20 and 602 seconds on the A100.
Given that the predictor’s execution time remains consistent across all models, it contributes less than 0.153\% of the total processing time on the L20 and 0.138\% on the A100, confirming that the overhead is negligible.

\subsubsection{Approach-2: Inter-batch Work Stealing}

\begin{figure}[h]
    \centering    \includegraphics[width=0.9\linewidth]{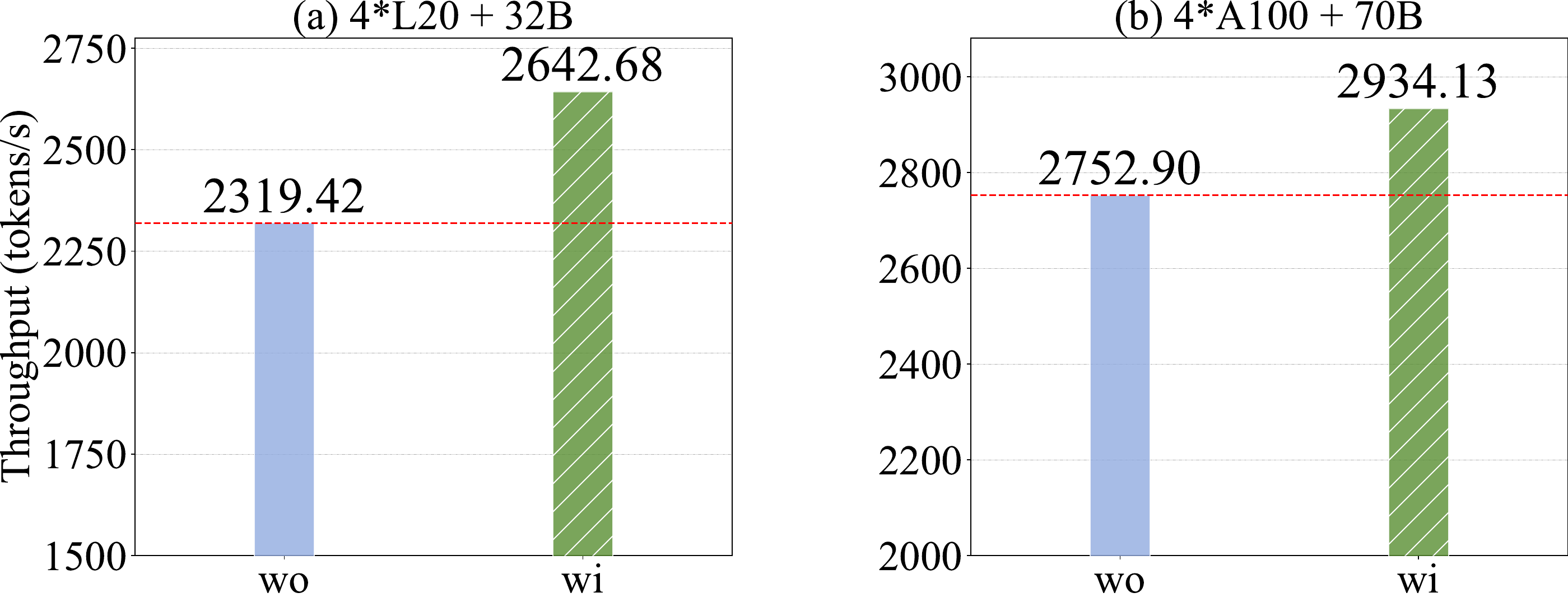}
    \vspace{-10pt}
    \caption{Ablation study for the decode batch load balance.}
    \label{fig:approach2}
\end{figure}

We report throughput results with (\textbf{wi}) or without (\textbf{wo}) the inter-batch work stealing approach in Figure~\ref{fig:approach2}.
The load balance scheduling at the timing for switching from prefill to decode is still kept and only the dynamic balance during the decode phase is removed.
L20 + 32B and A100 + 70B have 1.14$\times$ and 1.07$\times$ throughput improvement after activating our inter-batch work stealing approach, validating its effectiveness.

\subsubsection{Approach-3: Spatial Temporal Intensity Comparison}

\begin{figure}[h]
    \centering    \includegraphics[width=0.9\linewidth]{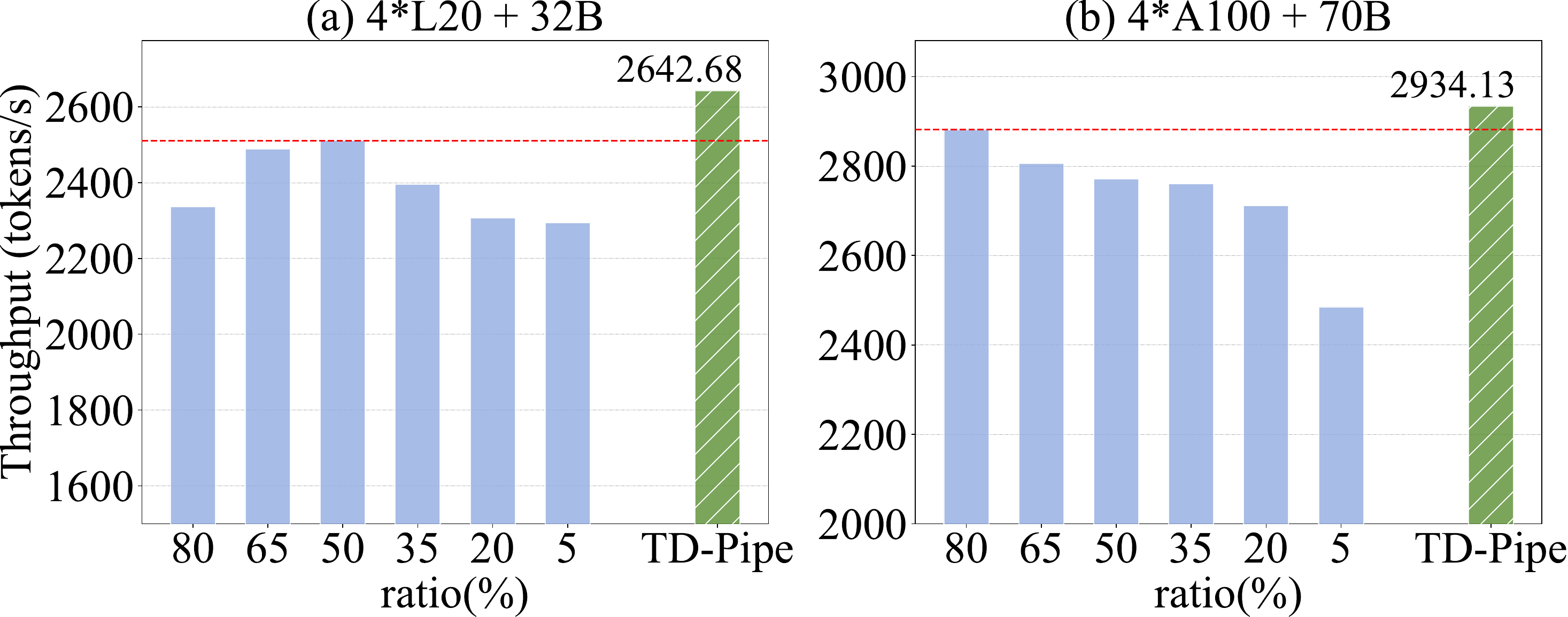}
    \vspace{-10pt}
    \caption{Ablation study for decode-to-prefill switching.}
    \label{fig:decode-to-prefill}
\end{figure}

Like evaluating the AI based greedy prefill approach, we introduce a hyperparameter called the request finish ratio to replace the spatial temporal intensity comparison approach.
For example, if this ratio is set to 50\%, the decode phase will switch to prefill once 50\% of the requests have completed.
In Figure~\ref{fig:decode-to-prefill}, the manually selected points perform well even when the ratio is somewhat extreme, due to the large memory capacity in these cases.
In comparison, the spatial temporal intensity comparison approach consistently achieves the highest throughput, illustrating its effectiveness.

\section{Related Work}

\textbf{LLM inference.} LLM inference has recently attracted significant research attention, resulting in a variety of systems designed to improve the efficiency and scalability of LLM inference. For instance, Orca~\cite{yu2022orca} introduces continuous batching to enhance throughput, while vLLM~\cite{vllm2024} proposes paged-attention for more granular KV cache management. Furthermore, both online and offline inference scenarios have been extensively studied. For online serving, DistServe~\cite{zhong2024distserve} proposes PD disaggregation to eliminate interference between the prefill and decode stages, and Sarathi~\cite{agrawal2024taming} implements a chunked prefill strategy to reduce contention between these operations. For offline inference, BlendServe~\cite{zhao2024blendserveoptimizingofflineinference} and BatchLLM ~\cite{zheng2025batchllmoptimizinglargebatched} leverage request reordering to accelerate processing for requests with shared prefixes and maximize resource utilization. 
Notably, TD-Pipe focuses on pipeline optimization to reduce bubbles, which is orthogonal to the request reordering method. \\
\textbf{Bubble Elimination for Pipeline Parallelism.} Pipeline parallelism is commonly employed in deep learning training and inference because it offers substantially lower communication overhead compared to tensor parallelism. However, complex dependency issues can give rise to pipeline bubbles, which ultimately lead to suboptimal device utilization. In terms of training, frameworks such as GPipe~\cite{huang2019gpipe}, Megatron ~\cite{narayanan2021efficient}, and Mobius~\cite{feng2023mobius} have introduced techniques to mitigate pipeline bubbles during the forward and backward passes. For inference, Sarathi~\cite{agrawal2024taming} initially proposed chunked prefill specifically to reduce pipeline bubbles.


\section{Conclusion}

To achieve the high-throughput LLM inference, especially on the bandwidth-constrained commodity hardware, we propose the \\temporally-disaggregated pipeline parallelism architecture.
Around this novel architecture, we identify key concerns and introduce the LLM inference system, TD-Pipe.
Basically, TD-Pipe disaggregates the prefill phase and decode phase in the temporal dimension.
To better exploit the architecture,TD-Pipe decouples the execution plane and the control plane for better device coordination.
In addition, TD-Pipe designs solutions for optimizing the prefill-to-decode and decode-to-prefill switching timing selection, and dynamically balancing the inter-batch workloads during the decode phase.
Experimental results demonstrate that TD-Pipe outperforms existing tensor parallel and pipeline parallel approaches by up to 1.91$\times$ and 2.73$\times$, respectively, when scaling to 4 GPUs on PCIe nodes.

\balance
\bibliographystyle{ACM-Reference-Format}
\bibliography{sample-base}

\end{document}